\documentclass[a4paper, 10pt, conference]{IEEEtran}
\IEEEoverridecommandlockouts                   

\usepackage{graphicx}
\usepackage{color}
\usepackage{amsmath}
\usepackage{multirow}
\usepackage{booktabs}
\usepackage{url}
\usepackage{amsmath}
\usepackage[position=t,singlelinecheck=off]{subfig}
\usepackage{adjustbox}
\usepackage{subfloat}
\usepackage[english]{babel}
\usepackage{comment}
\usepackage{tikz}
\usepackage{float}
\usetikzlibrary{positioning}

\newcommand\copyrighttext{%
  \footnotesize \textcopyright 2012 IEEE. Personal use of this material is permitted. Permission from IEEE must be obtained for all other uses, in any current or future media, including reprinting/republishing this material for advertising or promotional purposes, creating new collective works, for resale or redistribution to servers or lists, or reuse of any copyrighted component of this work in other works. This paper has been accepted for publication in the 2020 Twelfth International Conference on Quality of Multimedia Experience (QoMEX).}
\newcommand\copyrightnotice{%
\begin{tikzpicture}[remember picture,overlay]
\node[anchor=south,yshift=10pt] at (current page.south) {\fbox{\parbox{\dimexpr\textwidth-\fboxsep-\fboxrule\relax}{\copyrighttext}}};
\end{tikzpicture}%
}

\begin{document}

\title{Comparing emotional states induced by 360$^{\circ}$ videos via head-mounted display and computer screen}

\author{
 \IEEEauthorblockN{Jan-Niklas Voigt-Antons$^{1,2}$, Eero Lehtonen$^{3}$, Andres Pinilla Palacios$^1$, Danish Ali$^1$, Tanja Koji\'{c}$^1$, Sebastian M\"oller$^{1,2}$}
 \IEEEauthorblockA{$^1$Quality and Usability Lab, TU Berlin, Germany\\
 $^2$German Research Center for Artificial Intelligence (DFKI), Berlin, Germany\\
 $^3$Alto University, Finland}
}


\maketitle
\copyrightnotice

\begin{abstract}
In recent years 360$^{\circ}$ videos have been becoming more popular. For traditional media presentations, e.g., on a computer screen, a wide range of assessment methods are available. Different constructs, such as perceived quality or the induced emotional state of viewers, can be reliably assessed by subjective scales. Many of the subjective methods have only been validated using stimuli presented on a computer screen. This paper is using 360$^{\circ}$ videos to induce varying emotional states. Videos were presented 1) via a head-mounted display (HMD) and 2) via a traditional computer screen. Furthermore, participants were asked to rate their emotional state 1) in retrospect on the self-assessment manikin scale and 2) continuously on a 2-dimensional arousal-valence plane. In a repeated measures design, all participants (N = 18) used both presentation systems and both rating systems. Results indicate that there is a statistically significant difference in induced presence due to the presentation system. Furthermore, there was no statistically significant difference in ratings gathered with the two presentation systems. Finally, it was found that for arousal measures, a statistically significant difference could be found for the different rating methods, potentially indicating an underestimation of arousal ratings gathered in retrospect for screen presentation. In the future, rating methods such as a 2-dimensional arousal-valence plane could offer the advantage of enabling a reliable measurement of emotional states while being more embedded in the experience itself, enabling a more precise capturing of the emotional states.
\end{abstract}



\section{INTRODUCTION}
There is a steady increase in applications and research using virtual reality. In contrast to well-studied media types like audio, speech, and video, there are many factors related to the experience produced by virtual environments that have not been exhaustively studied. Understanding how people perceive the quality of VR systems is pivotal to improve the Quality of Experience (QoE) of users \cite{Le2012}. Given that QoE is a broad construct, there are many possible influencing factors associated with it. Some of these are of technical nature \cite{Reiter2014}, while others are related to the users' perception of the system \cite{Wechsung2014}. The present paper will be focused on the latter.

Multiple methods can be used to assess perception. Some of the most common are 1) direct measures (e.g. asking questions) \cite{Schatz2017}, 2) indirect measures (e.g. observing behavior) \cite{Robitza2016}, and 3) psychophysiological signals \cite{Antons2013}. Some of the direct approaches have been validated to be used in a paper version or on a screen. However, it is not always clear whether these methods can be used reliably within virtual environments. Recent studies suggest that it is possible to use questionnaires in virtual environments to assess constructs (i.e., presence) \cite{Regal2019}. However, it is not clear if the same applies to other constructs, such as emotional states.

Emotions are important in the field of QoE because they are pivotal in the overall experience of users. The methods previously mentioned (direct, indirect, and psychophysiological measures) can be used to estimate the emotional state of people. For example, to analyze how aroused someone is during and after using a product. Pictographic scales are often used to assess arousal (the intensity of an emotion) and valence (how negative or positive an emotion is) \cite{Bradley1994}. Usually, participants use these scales to rate their emotional state on a sheet of paper or on a computer screen. However, when VR is used for the presentation of stimuli, participants have to leave the virtual environment to give the rating and then enter again to proceed with the test. This breaks the entire experience and might bias the measures.

An additional problem of many methods used to assess the emotional state of participants is that the measure is taken after the actual experience (i.e., in retrospect). There are tools that allow taking measures of \emph{perceived quality} during the media stimulation (i.e., continuously) \cite{Perez2019}, but it is not clear if these tools work reliably for emotional states as well. Besides, the measurement method itself and the form of the media presentation can bias the results.

On the other hand, the spread of 360$^{\circ}$ videos has created an interesting source of media content for HMD/VR systems. Furthermore, it is known that immersive stimuli can lead to a stronger emotional response \cite{Fonseca2016}. However, it is not clear how to assess participants’ emotional states while watching a 360$^{\circ}$ video. To the knowledge of the authors, there is no research about the validity of a continuous rating method of emotional states for 360$^{\circ}$ videos, considering that the immersiveness of the stimuli is a potential influencing factor.

Therefore, the aim of this paper is 1) to investigate if evoked emotions change due to the used presentation system (HMD vs. computer screen), 2) analyze whether there are differences in retrospect vs. continuous measures, 3) to show if the two rating method result measures the underlying construct, and 4) understand how the immersiveness of the stimuli influences the emotional experience, as an overall effect and for both rating methods separately.

\subsection{Objectives}
In order to explore the possibility of gathering ratings for emotional responses in virtual reality, a study was designed, giving participants the opportunity to rate their emotional state. The focus of the study was on two main parameters: the presentation system and the rating method. There is already research in the area of immersive environments and emotional responses \cite{Marques2019} and how to gather emotional responses in a virtual environment \cite{Toet2019}. What has gotten less focus is how the difference in the immersive environment affects the emotional responses of the participant. To be able to gather ratings in a virtual environment, \cite{Toet2019} proposed to use an emotional grid to measure emotional responses. The comparison of the measured rating showed a high similarity to the measured values outside of HMD. The ratings were gathered in retrospect (i.e., after the stimulus presentation was over), showing the potential of using in HMD measures in general. Differentiation of conditions for the study has been created by using different text lengths (short title, medium, and longer paragraph), as well as different HMD devices with different screen resolutions. 

\begin{figure}[h!]
\centering
\includegraphics[width=0.3\textwidth]{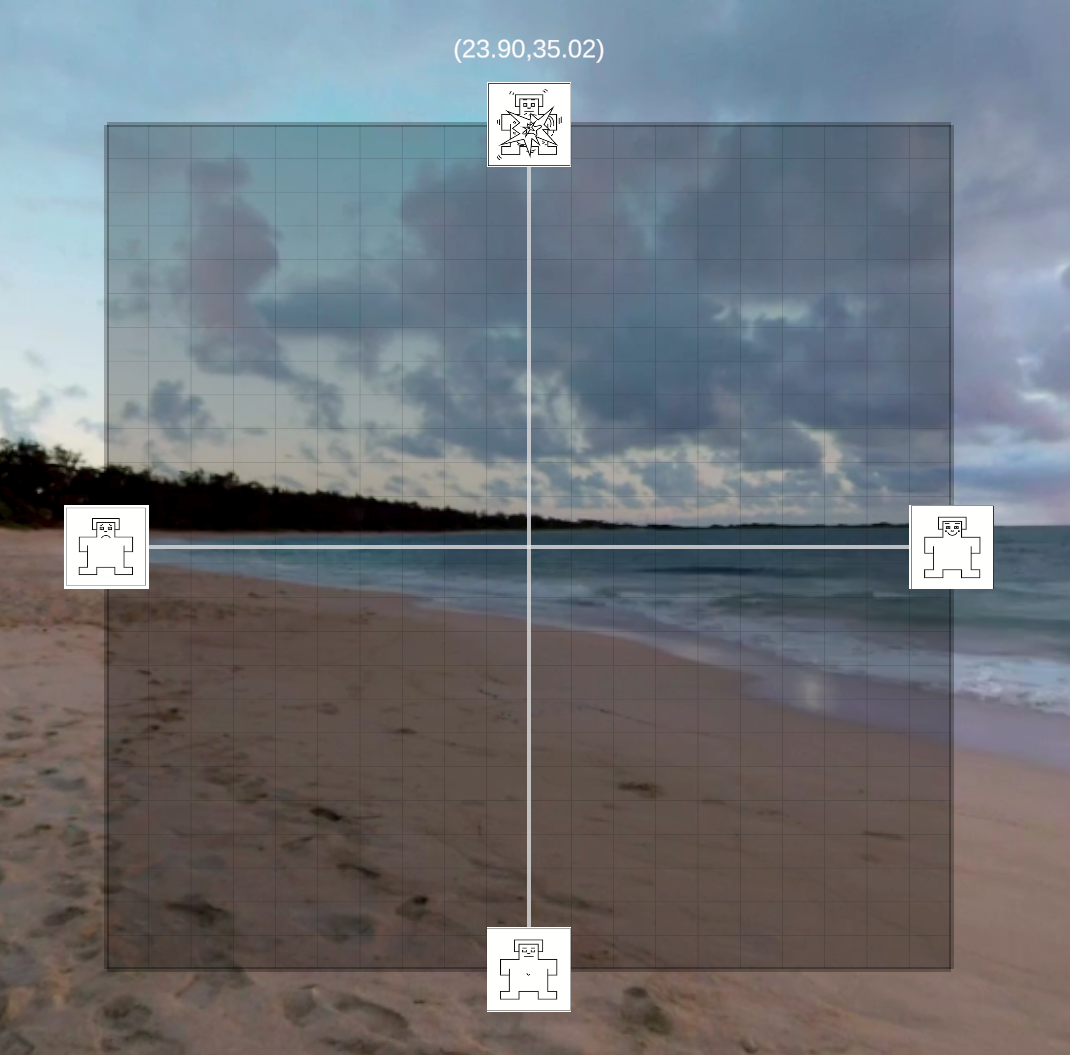}
\caption{Rating user interface of the continuous rating method. Shown with the background of one exemplary video.}
\label{fig:dmm}
\end{figure}

\begin{table*}[]
  \centering
  \caption{Used videos from database \cite{Li2017} in the experiment. To keep a constant duration for all stimuli, the selected time interval for each video is indicated. The quadrants or the orthogonal space with two dimensions: arousal and valence is stated (LALV=low arousal, low valence; LAHV=low arousal, high valence; HALV=high arousal, low valence; HAHV=high arousal, high valence).}
  \resizebox{2\columnwidth}{!}{
    \begin{tabular}{lllll}
    \toprule
\multicolumn{1}{l}{Video number} & \multicolumn{1}{l}{Name} & \multicolumn{1}{l}{Short link} & \multicolumn{1}{l}{Time interval}& \multicolumn{1}{l}{Quadrant}
    \\
    \midrule
    1 & The Displaced & https://youtu.be/ecavbpCuvkI & $2:23-3:23$ & LALV \\
    2 & Solitary Confinement & https://youtu.be/nDwulYcboDU & $0:00-1:00$ & LALV\\
    3 & Malaekahana Sunrise & https://youtu.be/-bIrUYM-GjU & $1:20-2:20$ & LAHV\\
    4 & Great Ocean Road & https://youtu.be/aszTdBlbfq0 & $0:00-1:00$ & LAHV\\
    5 & Jailbreak & https://youtu.be/vNLDRSdAj1U & $2:39-3:39$ & HALV\\
    6 & War Knows No Nation & https://youtu.be/CIbo0xLbNic & $4:44-5:44$ & HALV\\
    7 & Walk the Tight Rope & https://youtu.be/JtAzMFcUQ90 & $0:27 -1:27$ & HAHV\\
    8 & Puppies Host SourceFed For A Day & https://youtu.be/c7sA3EdXSUQ & $0:04-1:04$ & HAHV\\
    \bottomrule
    \end{tabular}
    }
  \label{tab:videos}
\end{table*}

The remaining sections of the present paper explain the methods and results of the study with the aim of answering these questions:
\begin{itemize}
    \item Were the selected stimuli able to evoke emotional responses (independent from presentations system and rating method)?
    \item Do the evoked emotions change due to the used presentation system (HMD vs. computer screen) (independently from the rating methods)?
    \item Do the two rating methods measure the same underlying construct?
    \item Does the continuous rating have an impact on the experienced presence (independently from the presentation system)? 
\end{itemize}

\section{RELATED WORK}
Two concepts are pivotal for the study of virtual environments: presence and immersion. Presence is defined as the sense of existing in a virtual environment \cite{Slater1993} or as the illusion of being in a real place \cite{Slater2009}. Early approaches defined immersion as a property of a system strictly related to its technical characteristics (i.e., the ability of the system to generate a realistic environment) \cite{Biocca1995}. In contrast, more recent approaches propose that immersion is also related to the psychological experience of the user \cite{Nilsson2016}. Therefore, the classical distinction between presence and immersion has become less clear.

Even though there is not a unified definition of presence, most authors agree that it is a dimensional construct. According to \cite{Heeter1992}, it consists of three dimensions: personal presence (simulation of real-world stimuli in the virtual world), social presence (existence of other people in the virtual world), and environmental presence (the ability of the virtual world to adapt itself to the user). This definition is coherent with other findings, suggesting that presence is strongly related to the subjective experience of the user \cite{Nilsson2016}.

The lack of a unified definition of presence makes it more complicated to measure it. However, \cite{Schubert2001} used a factor analysis to enable a standardized measurement of perceived presence with a subjective scale: the Igroup Presence Questionnaire (IPQ). Other commonly used questionnaires are the Slater-Usoh-Steed (SUS) Questionnaire \cite{ Slater1994} and the Presence Questionnaire (PQ) \cite{Witmer1998}.

Previous research suggests that there is a correlation with immersion and presence, as well as some suggestions on the correlation between immersion and emotional states \cite{Banos2004}.  Further findings show that emotional states influence the sense of presence in virtual environments \cite{Ling2013} and that more immersive environments produce stronger emotional responses \cite{Marques2019}.  Other authors have investigated the impact of perception and presence on emotional reactions \cite{Diemer2015} and how the stereoscopy (depth and 3D) influence presence and emotions \cite{Banos2008}. What is still widely unknown is in which conditions immersion affects emotions and which emotional dimensions it affects. 

The assessment of emotional states also depends on the theoretical approach used by the researcher. Some authors propose that emotions are better described in terms of categorical variables \cite{Ekman1971}, while others opt for the usage of continuous variables. An iconic example of the latter can be found in the Circumplex Model of Affect \cite{Russell1980}, which consists of an orthogonal space with two dimensions: arousal and valence. Consequently, there are instruments that allow measuring emotional responses in terms of categories, such as Pick A Mood (PAM) \cite{Desmet2016}, or in terms of dimensional variables, such as the Self-Assessment Manikin (SAM) \cite{Bradley1994}. The former allows assessing emotional states as a discrete selection, while the latter consists of three-dimensional pictographic scales: arousal, valence, and dominance.

Several instruments that can be used to evoke emotions. For example, the International Affective Digitized Sounds (IADS) \cite{Bradley2007}, the International Affective Pictures System (IAPS) \cite{Lang2005}, a battery of films for emotion elicitation \cite{Gross1995}, and a public database of 360$^{\circ}$ videos \cite{Li2017}. Given that more immersive environments tend to elicit more intense emotional responses \cite{ Visch2010}, it is likely that emotion elicitation via HMD content is more effective than via traditional mediums.

To be able to gather ratings in a virtual environment, \cite{Toet2019} proposed to use an emotional grid to measure emotional responses. The comparison of the measured rating showed a high similarity to the measured values outside of the virtual environment. Although ratings were gathered in retrospect, after the stimulus presentation was over, this shows the potential of using measures in a virtual environment in general. 

\section{METHODS}
\subsection{Participants}Eighteen persons participated in the study. Their age was between 20 and 46 years old ({\textit{M}} = 29.2; {\textit{SD}} = 7.55). Seven were women, and 11 were men. No participant reported impairments concerning hearing or visual accuracy. All participants provided written informed consent before participating in the experiment.

\subsection{Stimuli}
A set of eight 360$^{\circ}$ videos were used. Each of the videos had a duration of 60 seconds (see Table \ref{tab:videos}). The videos were selected so that two videos from each quadrant were included. The quadrants were the following:

\begin{itemize}
\item High arousal, high valence
\item High arousal, low valence
\item Low arousal, high valence
\item Low arousal, low valence
\end{itemize}

The videos were taken from the database described in \cite{Li2017}. The database was created to provide a selection of 360$^{\circ}$ videos that are able to evoke specific emotions. 
To be able to compensate for the different duration of the videos provided in the database, three experts identified 60 seconds intervals of each video that should evoke the intended emotion in the viewers (two psychologists and one computer scientist). The duration of stimuli was selected to keep the duration of stimuli constant while keeping the overall duration of the experiment as short as possible. 

\subsection{Conditions}
A 2 x 2 within-subjects factor design was used, with the type of stimulus presentation as the first within-subjects factor (HMD vs. computer screen) and type of evaluation (continuous vs. retrospective) as the second within-subjects factor. The same videos were used for each condition. Therefore, each participant saw each video four times. The order of the conditions and the videos within a condition was randomly selected for each participant.

\subsection{Apparatus}
The stimuli were presented using a virtual environment that was developed in Unity. An Oculus Quest (HMD condition) and an MSI gaming laptop (screen condition) were used. The screen size of the gaming laptop was 44cm. The approximate distance between the participant and the screen in the screen condition was 60 centimeters. As the videos shown to the participants were 360$^{\circ}$ videos, the turning of the camera angle happened with arrow keys (screen condition) and with head movements (HMD condition). Participants evaluated the videos using a controller (HMD condition) and a mouse (screen condition).

\subsection{Rating scales}
The two different rating systems used in the experiment were retrospective and continuous evaluation.

In the retrospective evaluation, the participants answered three questions after each video on a separate desktop. When doing the retrospective evaluation with the HMD condition, the participant had to take the headset off to make the evaluation. The questions were the following:

\begin{itemize}
\item In the computer-generated world I had a sense of "being there" (scale 1 to 5), Question G1 from \cite{Schubert2001} only one item was selected to keep the stimulus duration as short as possible
\item Valence rating (scale 1 to 9) taken from \cite{Bradley1994}
\item Arousal rating (scale 1 to 9) taken from \cite{Bradley1994}
\end{itemize}

After answering the questions, the participants proceeded to the next video. 

In the continuous evaluation, the participants gave subjective ratings regarding their emotional state while watching the videos. While watching the video, the participant could evaluate it as many times as they felt necessary from their own incentive. Participants evaluated each video by clicking on a point in a two-dimensional orthogonal grid that represented the valence and arousal dimensions of the Self-Assessment Manikin (SAM) [2] (see Figure 1). The grid was stationary in the virtual environment. To get the final valence and arousal value for each participant, apparatus, video combination in the continuous rating system, we averaged all the ratings given by each participant in the given condition during the 60 second time window of the stimulus.

\subsection{Procedure}
The experiment began by giving the participant an introduction sheet regarding the experiment. The introduction sheet explained the general information of the experiment, important concepts (arousal, valence), the rating methods (continuous, retrospective), and the outline of the experiment. After having informed the participants about the experiment, they read and signed the informed consent as well as the demographics questionnaire. After that, participants tested the usage of the continuous evaluation plane by watching one video with a desktop before proceeding to the real experiment. In the experiment, as the order of conditions was randomized across participants, half of the participants saw the videos first in HMD and then on screen. The other half saw them first on-screen and then in HMD. Similarly, the order of the type of evaluation was randomized. Consequently, half of the participants did the continuous evaluation first, and the other half did the retrospective evaluation first. Participants were able to freely move, turn, and lock around during video playback. After finishing all of the four conditions, the participants filled the final questionnaire, which concluded the experiment.

\begin{table}[htbp]
  \centering
  \caption{Statistically significant effects of \textit{video} (eight videos), \textit{presentation system} (HMD vs. screen) and \textit{rating method} (retrospect vs. continuous) on experiences emotion (valence and arousal scale) and perceived presence.}
  \resizebox{\columnwidth}{!}{
    \begin{tabular}{lllcllrl}
    \toprule
\multicolumn{1}{l}{Effect} & \multicolumn{1}{l}{Parameter} & \multicolumn{1}{c}{$df_{\textnormal n}$} & \multicolumn{1}{l}{$df_{\textnormal d}$} & \multicolumn{1}{c}{$F$} & \multicolumn{1}{c}{$p$} & \multicolumn{1}{c}{$\eta_{\textnormal G}^2$}
    \\
    \midrule
    Video & SAM\_A & $7$     & $13$    & $23.76$ & $ <.001$ & $0.65$ \\
    Video & SAM\_V &  $7$     & $13$    & $46.99$ & $ <.001$ & $0.78$ \\
    Presentation system & Presence &  $1$     & $22$    & $6.14$ & $ .022$ & $0.22$ \\
    Rating method & SAM\_A &  $1$     & $13$    & $5.58$ & $ .034$ & $0.30$ \\
    \bottomrule
    \end{tabular}%
    }
  \label{tab:anova_results}%
\end{table}%

\subsection{Ethics}
The experiment was approved by the local ethics committee of the Faculty IV of the Technische Universität Berlin (approval number FT-2019-05).
The experimental procedure did not represent any risk for human health. The emotional effect produced by the videos did not have any long term effects.

\section{RESULTS}
A repeated measures Analysis of Variance (ANOVA) was performed to analyze whether the  \textit{presentation system} (screen vs. HMD) and \textit{rating method} (retrospect vs. continuous) had an effect on the evaluation of the videos. A summary of all significant effects that will be explained in the following sections is given in Table \ref{tab:anova_results}.

\begin{figure}[]
\centering
\includegraphics[width=0.5\textwidth]{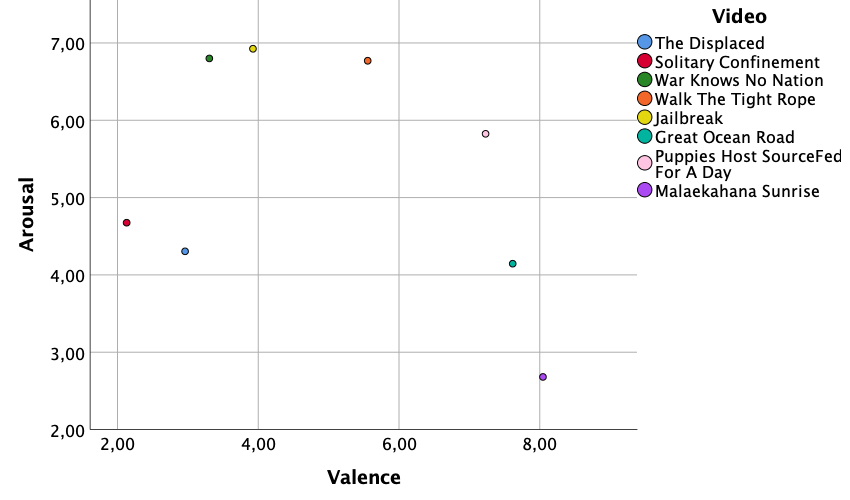}
\caption{Average values for SAM ratings for the dimensions arousal (y-axis) and valence (x-axis) over all participants.}
\label{fig:scatterVA}
\end{figure}

\subsection{Stimulus selection}
As shown in Figure \ref{fig:scatterVA}, there was a statistically significant influence of the independent variable \textit{video} on the \textit{arousal} and \textit{valence} ratings (see Table \ref{tab:anova_results} for test statistics). There are two videos in each of the four quadrants.

\subsection{Influence of presentation system on presence}
\begin{figure}[h!]
\centering
\includegraphics[width=0.5\textwidth]{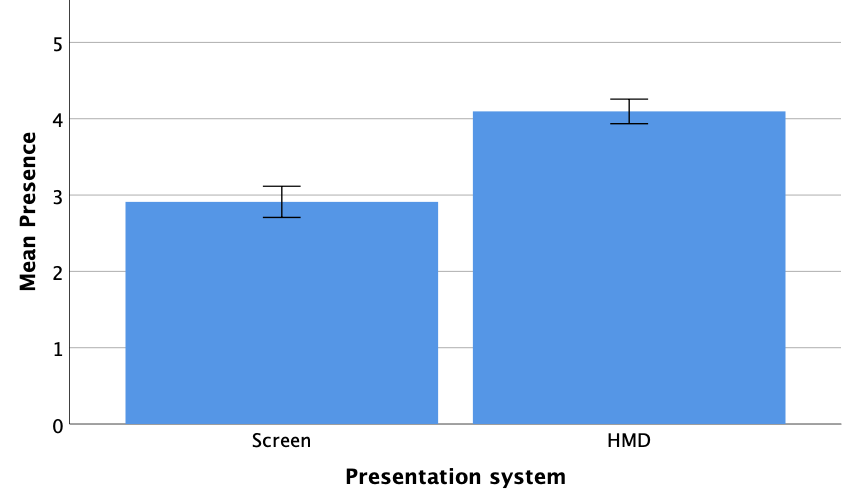}
\caption{Average values for presence ratings for the two presentation systems (screen vs. HMD) over all participants. Whiskers denote the 95\% confidence interval.}
\label{fig:presence}
\end{figure}

The \textit{presentation system} had a significant influence on the sense of \textit{presence} (see Table \ref{tab:anova_results} for test statistics). As shown in Figure \ref{fig:presence}, the average value over all participants for the presentation system screen (M=2.85, SE=0.20) is lower compared to HMD (M=4.05, SE=0.16).

\subsection{Influence of presentation system on valence}
\begin{figure}[H]
\centering
\includegraphics[width=0.5\textwidth]{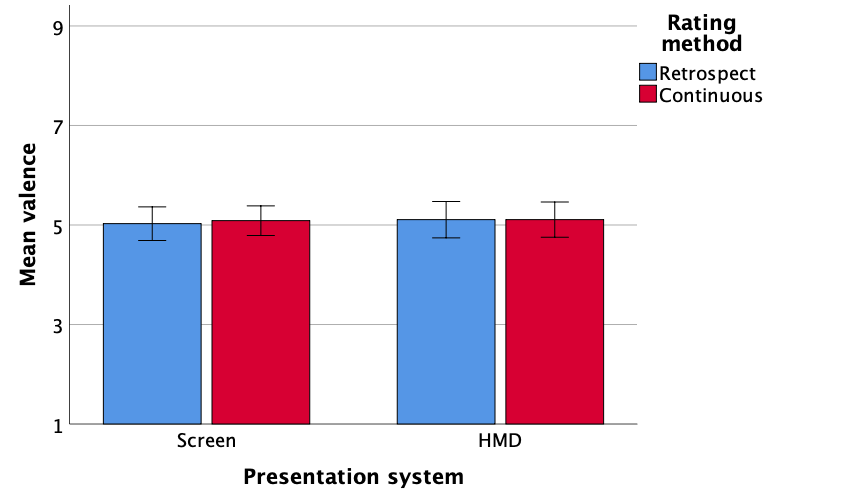}
\caption{Average values for valence ratings for the two presentation systems (screen vs. HMD) and the two rating methods (retrospect vs. continuous) over all participants. Whiskers denote the 95\% confidence interval.}
\label{fig:valence}
\end{figure}

The \textit{presentation system} had no significant influence on \textit{valence}. As Figure \ref{fig:valence} suggests, the average value over all participants for the screen condition is similar to HMD for both rating methods. The mean of retrospective and continuous rating in the screen condition were 5.03 (SE=0.16) and 5.09 (SE=0.14), respectively. In the HMD condition, the mean retrospective rating was 5.11 (SE=0.17) and 5.11 (SE=0.16), respectively.

\subsection{Influence of rating methods on arousal}
\begin{figure}[H]
\centering
\includegraphics[width=0.5\textwidth]{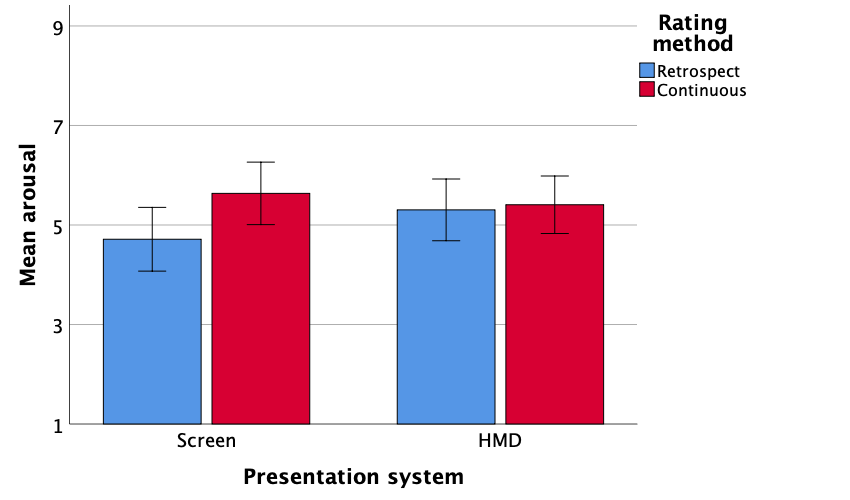}
\caption{Average values for arousal ratings for the two presentation systems (screen vs. HMD) and the two rating methods (retrospect vs. continuous) over all participants. Whiskers denote the 95\% confidence interval.}
\label{fig:arousal}
\end{figure}

\textit{Rating methods} had a statistically significant influence on \textit{arousal} (see Table \ref{tab:anova_results} for test statistics). As shown in Figure \ref{fig:arousal}, the average arousal value for the presentation system HMD over all participants for the rating methods retrospectively (M=5.30, SE=0.29) is similar to the continuous rating method (M=5.41, SE=0.27). For the presentation method screen, we have a lower arousal rating for the retrospective rating method (M=4.70, SE=0.30), compared to the continuous rating method (M=5.64, SE=0.29).

\subsection{Intra-class correlations for arousal and valence}
The agreement of the used rating methods (retrospect vs. continuous) regarding the ratings across participants was assessed by the inter-rater reliability (IRR) using a two-way mixed, absolute agreement, average-measures intra-class correlation (ICC) \cite{hallgren2012}. The average resulting ICCs regarding the eight videos suggest excellent reliability \cite{cicchetti1994} for the valence score, total average ICC = 0.80, p $<$ 0.05, and of good reliability \cite{cicchetti1994} for the arousal score, total average ICC = 0.673, p $<$ 0.05, indicating that (1) the rating methods had a high degree of agreement and (2) that valence and arousal were rated similarly across the rating methods.

\section{DISCUSSION}
\subsection{Stimulus selection}
From the results, we can see that the used stimuli were able to create the desired emotional responses in the participants. This can be seen in Figure 2, where the average emotional responses over each condition have been calculated for each video. Each quadrant from the origin (5,5) contains two videos as expected.

\subsection{Influence of presentation system on presence}
Regarding presence felt in the virtual world, we were able to repeat the results recorded from previous studies, e.g., \cite{Marques2019}, which indicates a greater feeling of presence in the HMD condition than in the screen condition. 

\subsection{Influence of presentation system on valence}
When comparing the impact of the presentation system and the rating methods has on valence, no significant difference between the average value over ratings could be found. This could indicate that neither the immersiveness, which was implemented in our study by using HMD in contrast to a screen representation, nor the rating methods had a statistically significant effect on participants' emotional reactions. 

\subsection{Influence of rating methods on arousal}
On the other hand, when comparing the impact of the presentation system and the rating methods have on arousal, statically significant differences can be found. For the HMD condition, the two rating methods resulted in similar average values for arousal. For the presentations method screen, on the other hand, the continuous rating system resulted in statistically significant higher average values than the retrospective rating system. This effect could be addressed to the fact that participants were rating their emotional arousal in retrospect. Therefore, they might have underestimated the state in which they were during the stimulus presentation. It could also be that ratings are less accurate in the continuous rating condition because participants have to pay attention to the stimulus and to the rating tool at the same time.

\subsection{Intra-class correlations for arousal and valence}
The obtained intra-class correlations for arousal and valence indicate that both rating methods were rated similarly across the rating methods. We assume that also the continuous rating methods could prove as a valid measure in the assessment of emotional states. Based on the good to excellent intra-class correlations, we assume that different values for the two rating methods indicate the real state of participants and therefore were correctly obtained.

\section{CONCLUSION}
It was shown that the selected 360$^{\circ}$ video stimuli could successfully be used to evoke the intended emotional reactions in the participants. Furthermore, evidence is provided suggesting that the feeling of presence was higher for the most immersive presentation system (HMD). The two different rating methods were able to capture the same construct (perceived emotion as measured on the dimensions arousal and valence) and therefore offer a real-time assessment of the emotional states of participants. A continuous rating task of emotional responses is useful for other experiments, especially for stimuli with a long duration, allowing a more natural rating procedure within the stimulus environment itself. Future research could help to study further the temporal effects of continuous ratings. For example, in traditional video, audio, or speech quality experiments. Another possibility could be to make only the rating method visible for the participants when they press a button. Then the scale could directly be in focus and quickly rated.

\bibliographystyle{IEEEtran}
\bibliography{references.bib} 

\begin{thebibliography}{10}
\providecommand{\url}[1]{#1}
\csname url@samestyle\endcsname
\providecommand{\newblock}{\relax}
\providecommand{\bibinfo}[2]{#2}
\providecommand{\BIBentrySTDinterwordspacing}{\spaceskip=0pt\relax}
\providecommand{\BIBentryALTinterwordstretchfactor}{4}
\providecommand{\BIBentryALTinterwordspacing}{\spaceskip=\fontdimen2\font plus
\BIBentryALTinterwordstretchfactor\fontdimen3\font minus
  \fontdimen4\font\relax}
\providecommand{\BIBforeignlanguage}[2]{{%
\expandafter\ifx\csname l@#1\endcsname\relax
\typeout{** WARNING: IEEEtran.bst: No hyphenation pattern has been}%
\typeout{** loaded for the language `#1'. Using the pattern for}%
\typeout{** the default language instead.}%
\else
\language=\csname l@#1\endcsname
\fi
#2}}
\providecommand{\BIBdecl}{\relax}
\BIBdecl

\bibitem{Le2012}
P.~Le~Callet, S.~M{\"o}ller, A.~Perkis \emph{et~al.}, ``Qualinet white paper on
  definitions of quality of experience,'' \emph{European network on quality of
  experience in multimedia systems and services (COST Action IC 1003)}, vol.~3,
  no. 2012, 2012.

\bibitem{Reiter2014}
U.~Reiter, K.~Brunnstr{\"o}m, K.~De~Moor, M.-C. Larabi, M.~Pereira,
  A.~Pinheiro, J.~You, and A.~Zgank, ``Factors influencing quality of
  experience,'' in \emph{Quality of experience}.\hskip 1em plus 0.5em minus
  0.4em\relax Springer, 2014, pp. 55--72.

\bibitem{Wechsung2014}
\BIBentryALTinterwordspacing
I.~Wechsung and K.~De~Moor, \emph{Quality of Experience Versus User
  Experience}.\hskip 1em plus 0.5em minus 0.4em\relax Cham: Springer
  International Publishing, 2014, pp. 35--54. [Online]. Available:
  \url{https://doi.org/10.1007/978-3-319-02681-7_3}
\BIBentrySTDinterwordspacing

\bibitem{Schatz2017}
R.~{Schatz}, A.~{Sackl}, C.~{Timmerer}, and B.~{Gardlo}, ``Towards subjective
  quality of experience assessment for omnidirectional video streaming,'' in
  \emph{2017 Ninth International Conference on Quality of Multimedia Experience
  (QoMEX)}, May 2017, pp. 1--6.

\bibitem{Robitza2016}
W.~{Robitza} and A.~{Raake}, ``(re-)actions speak louder than words? a novel
  test method for tracking user behavior in web video services,'' in \emph{2016
  Eighth International Conference on Quality of Multimedia Experience (QoMEX)},
  June 2016, pp. 1--6.

\bibitem{Antons2013}
J.~{Antons}, F.~{Köster}, S.~{Arndt}, S.~{Möller}, and R.~{Schleicher},
  ``Changes of vigilance caused by varying bit rate conditions,'' in \emph{2013
  Fifth International Workshop on Quality of Multimedia Experience (QoMEX)},
  July 2013, pp. 148--151.

\bibitem{Regal2019}
G.~Regal, J.-N. Voigt-Antons, S.~Schmidt, J.~Schrammel, T.~Kojić,
  M.~Tscheligi, and S.~Möller, ``Questionnaires embedded in virtual
  environments: reliability and positioning of rating scales in virtual
  environments,'' \emph{Quality and User Experience}, vol.~4, no.~1, pp. 1--13,
  oct 2019.

\bibitem{Bradley1994}
M.~M. Bradley and P.~J. Lang, ``Measuring emotion: the self-assessment manikin
  and the semantic differential,'' \emph{Journal of behavior therapy and
  experimental psychiatry}, vol.~25, no.~1, pp. 49--59, 1994.

\bibitem{Perez2019}
P.~{Pérez} and J.~{Escobar}, ``Miro360: A tool for subjective assessment of
  360 degree video for itu-t p.360-vr,'' in \emph{2019 Eleventh International
  Conference on Quality of Multimedia Experience (QoMEX)}, June 2019, pp. 1--3.

\bibitem{Fonseca2016}
D.~Fonseca and M.~Kraus, ``A comparison of head-mounted and hand-held displays
  for 360 videos with focus on attitude and behavior change,'' in
  \emph{Proceedings of the 20th International Academic Mindtrek Conference},
  2016, pp. 287--296.

\bibitem{Marques2019}
T.~Marques, M.~Vairinhos, and P.~Almeida, ``How vr 360{\textordmasculine}
  impacts the immersion of the viewer of suspense av content,'' in
  \emph{Proceedings of the 2019 ACM International Conference on Interactive
  Experiences for TV and Online Video}, 2019, pp. 239--246.

\bibitem{Toet2019}
A.~Toet, F.~Heijn, A.-M. Brouwer, T.~Mioch, and J.~B. van Erp, ``The emojigrid
  as an immersive self-report tool for the affective assessment of 360 vr
  videos,'' in \emph{International Conference on Virtual Reality and Augmented
  Reality}.\hskip 1em plus 0.5em minus 0.4em\relax Springer, 2019, pp.
  330--335.

\bibitem{Li2017}
B.~J. Li, J.~N. Bailenson, A.~Pines, W.~J. Greenleaf, and L.~M. Williams, ``A
  public database of immersive vr videos with corresponding ratings of arousal,
  valence, and correlations between head movements and self report measures,''
  \emph{Frontiers in Psychology}, vol.~8, p. 2116, 2017.

\bibitem{Slater1993}
M.~Slater and M.~Usoh, ``Representations systems, perceptual position, and
  presence in immersive virtual environments,'' \emph{Presence: Teleoperators
  \& Virtual Environments}, vol.~2, no.~3, pp. 221--233, 1993.

\bibitem{Slater2009}
M.~Slater, ``Place illusion and plausibility can lead to realistic behaviour in
  immersive virtual environments,'' \emph{Philosophical Transactions of the
  Royal Society B: Biological Sciences}, vol. 364, no. 1535, pp. 3549--3557,
  2009.

\bibitem{Biocca1995}
F.~Biocca and B.~Delaney, ``Immersive virtual reality technology,''
  \emph{Communication in the age of virtual reality}, vol.~15, p.~32, 1995.

\bibitem{Nilsson2016}
N.~C. Nilsson, R.~Nordahl, and S.~Serafin, ``Immersion revisited: A review of
  existing definitions of immersion and their relation to different theories of
  presence.'' \emph{Human Technology}, vol.~12, no.~2, 2016.

\bibitem{Heeter1992}
\BIBentryALTinterwordspacing
C.~Heeter, ``\BIBforeignlanguage{en}{Being {There}: {The} {Subjective}
  {Experience} of {Presence}},'' \emph{\BIBforeignlanguage{en}{Presence:
  Teleoperators and Virtual Environments}}, vol.~1, no.~2, pp. 262--271, Jan.
  1992. [Online]. Available:
  \url{http://www.mitpressjournals.org/doi/10.1162/pres.1992.1.2.262}
\BIBentrySTDinterwordspacing

\bibitem{Schubert2001}
T.~Schubert, F.~Friedmann, and H.~Regenbrecht, ``The experience of presence:
  Factor analytic insights,'' \emph{Presence: Teleoperators \& Virtual
  Environments}, vol.~10, no.~3, pp. 266--281, 2001.

\bibitem{Slater1994}
M.~Slater, M.~Usoh, and A.~Steed, ``\BIBforeignlanguage{en}{Depth of {Presence}
  in {Virtual} {Environments}},'' \emph{\BIBforeignlanguage{en}{Presence:
  Teleoperators and Virtual Environments}}, vol.~3, no.~2, pp. 130--144, Jan.
  1994.

\bibitem{Witmer1998}
B.~G. Witmer and M.~J. Singer, ``\BIBforeignlanguage{en}{Measuring {Presence}
  in {Virtual} {Environments}: {A} {Presence} {Questionnaire}},''
  \emph{\BIBforeignlanguage{en}{Presence: Teleoperators and Virtual
  Environments}}, vol.~7, no.~3, pp. 225--240, Jun. 1998.

\bibitem{Banos2004}
R.~M. Ba{\~n}os, C.~Botella, M.~Alca{\~n}iz, V.~Lia{\~n}o, B.~Guerrero, and
  B.~Rey, ``Immersion and emotion: their impact on the sense of presence,''
  \emph{Cyberpsychology \& behavior}, vol.~7, no.~6, pp. 734--741, 2004.

\bibitem{Ling2013}
Y.~Ling, H.~T. Nefs, W.-P. Brinkman, C.~Qu, and I.~Heynderickx,
  ``\BIBforeignlanguage{en}{The relationship between individual characteristics
  and experienced presence},'' \emph{\BIBforeignlanguage{en}{Computers in Human
  Behavior}}, vol.~29, no.~4, pp. 1519--1530, Jul. 2013.

\bibitem{Diemer2015}
J.~Diemer, G.~W. Alpers, H.~M. Peperkorn, Y.~Shiban, and A.~Mühlberger, ``The
  impact of perception and presence on emotional reactions: a review of
  research in virtual reality,'' \emph{Frontiers in Psychology}, vol.~6, p.~26,
  2015.

\bibitem{Banos2008}
R.~M. Ba{\~n}os, C.~Botella, I.~Rubi{\'o}, S.~Quero, A.~Garc{\'\i}a-Palacios,
  and M.~Alca{\~n}iz, ``Presence and emotions in virtual environments: The
  influence of stereoscopy,'' \emph{CyberPsychology \& Behavior}, vol.~11,
  no.~1, pp. 1--8, 2008.

\bibitem{Ekman1971}
\BIBentryALTinterwordspacing
P.~Ekman and W.~V. Friesen, ``\BIBforeignlanguage{en}{Constants across cultures
  in the face and emotion.}'' \emph{\BIBforeignlanguage{en}{Journal of
  Personality and Social Psychology}}, vol.~17, no.~2, pp. 124--129, 1971.
  [Online]. Available: \url{http://doi.apa.org/getdoi.cfm?doi=10.1037/h0030377}
\BIBentrySTDinterwordspacing

\bibitem{Russell1980}
J.~A. Russell, ``A circumplex model of affect.'' \emph{Journal of Personality
  and Social Psychology}, vol.~39, no.~6, pp. 1161--1178, 1980.

\bibitem{Desmet2016}
N.~R. P.M.A.~Desmet, M.H.~Vastenburg, ``Mood measurement with pick-a-mood:
  review of current methods and design of a pictorial self-reportscale,''
  \emph{J. Design Research}, vol.~14, no.~3, pp. 241 -- 279, 2016.

\bibitem{Bradley2007}
M.~M. Bradley and P.~J. Lang, ``The international affective digitized sounds (;
  iads-2): Affective ratings of sounds and instruction manual,''
  \emph{University of Florida, Gainesville, FL, Tech. Rep. B-3}, 2007.

\bibitem{Lang2005}
P.~J. Lang, ``International affective picture system (iaps): Affective ratings
  of pictures and instruction manual,'' \emph{Technical report}, 2005.

\bibitem{Gross1995}
J.~J. Gross and R.~W. Levenson, ``Emotion elicitation using films,''
  \emph{Cognition \& emotion}, vol.~9, no.~1, pp. 87--108, 1995.

\bibitem{Visch2010}
V.~T. Visch, E.~S. Tan, and D.~Molenaar, ``The emotional and cognitive effect
  of immersion in film viewing,'' \emph{Cognition and Emotion}, vol.~24, no.~8,
  pp. 1439--1445, 2010.

\bibitem{hallgren2012}
K.~A. Hallgren, ``Computing inter-rater reliability for observational data: An
  overview and tutorial,'' \emph{Tutorials in Quantitative Methods for
  Psychology}, vol.~8, no.~1, pp. 23--34, feb 2012.

\bibitem{cicchetti1994}
D.~V. Cicchetti, ``Guidelines, criteria, and rules of thumb for evaluating
  normed and standardized assessment instruments in psychology.''
  \emph{Psychological Assessment}, vol.~6, no.~4, pp. 284--290, dec 1994.

\end{thebibliography}

\end{document}